\documentclass[aps,prd,twocolumn,showpacs,preprintnumbers,amsmath,amssymb]{revtex4}

\usepackage{amsmath,amssymb}
\usepackage{color}
\usepackage[dvips]{graphicx}

\def\apj{ApJ}%
\def\aap{A\&A}%
\def\jcap{J. Cosmology Astropart. Phys.}%
\def\mnras{MNRAS}%
\def\prd{Phys.~Rev.~D}%
\begin{document}

\title{Decaying dark matter: a stacking analysis of galaxy clusters\\ to
improve on current limits}

\author{C. Combet}
\email{celine.combet@lpsc.in2p3.fr}
\affiliation{Laboratoire de Physique Subatomique et de Cosmologie,
      Universit\'e Joseph Fourier Grenoble 1/CNRS/IN2P3/INPG,
      53 avenue des Martyrs, 38026 Grenoble, France}

\author{D. Maurin}
\affiliation{Laboratoire de Physique Subatomique et de Cosmologie,
      Universit\'e Joseph Fourier Grenoble 1/CNRS/IN2P3/INPG,
      53 avenue des Martyrs, 38026 Grenoble, France}

\author{E. Nezri}
\affiliation{Laboratoire d'Astrophysique de Marseille - LAM, Universit\'e 
d'Aix-Marseille \& CNRS, UMR7326, 38 rue F. Joliot-Curie, 13388 
Marseille Cedex 13, France}

\author{E. Pointecouteau}
\affiliation{(1) Universit\'e de Toulouse (UPS-OMP), Institut de Recherche en Astrophysique et Planetologie\\
		(2) CNRS, UMR 5277, 9 Av. colonel Roche, BP 44346, F 31028 Toulouse cedex 4, France}

\author{J.~A. Hinton}
\affiliation{Dept. of Physics and Astronomy, University of Leicester, Leicester, LE1 7RH, UK}

\author{R. White}
\affiliation{Dept. of Physics and Astronomy, University of Leicester, Leicester, LE1 7RH, UK}

\date{\today}

\begin{abstract} 

  We show that a stacking approach to galaxy clusters can improve
  current limits on decaying dark matter by a factor $\gtrsim 5-100$,
  with respect to a single source analysis, for all-sky instruments
  such as Fermi-LAT. Based on the largest sample of X-ray-selected
  galaxy clusters available to date (the MCXC meta-catalogue), we
  provide all the astrophysical information, in particular the
  astrophysical term for decaying dark matter, required to perform an
  analysis with current instruments.

\end{abstract}


\maketitle

\section{Introduction}
The nature of dark matter (DM) is one of the major questions in both
modern astrophysics and fundamental physics. In the most popular
extensions of the standard model, an explicit symmetry provides stable
annihilating DM candidates. Nevertheless, decaying DM is an
equally well-motivated scenario; it arises in several consistent
extensions beyond the standard model. Among those candidates and
frameworks, a non-exhaustive list includes: gravitinos in R-parity
breaking models~\cite{2000PhLB..485..388T,2007JHEP...03..037B},
sterile neutrinos~\cite{2005PhLB..631..151A}, hidden sector gauge
bosons~\cite{2009PhLB..671...71C}, hidden sector
particles~\cite{2010JCAP...03..024A}, right-handed
neutrinos/sneutrinos~\cite{2009JHEP...04..044P}, bound states of
strongly interacting particles~\cite{2007PhLB..654..110H}. To be
viable in the DM context, these models should induce a DM lifetime
longer than the age of the universe.

For the case considered here (see below), the total flux expected from
decay, in a given direction $(l,b)$ (Galactic coordinates) and integrated
over the solid angle $\Delta\Omega$, is given by the product of a particle
physics term with an astrophysical term $D$
\begin{equation}
  \frac{d\phi(E,l,b,\Delta\Omega)}{dE} = \frac{dN}{dE}(E) \times D(l,b,\Delta\Omega)\,.
\end{equation}
Depending on the decaying DM candidate, the photon spectrum $dN/dE$
can be a mono-energetic line (from $\gamma\gamma$ and $\gamma\nu$
channels), or a continuum (e.g., for some gravitino candidates above a
few hundred GeV \cite{2011arXiv1110.1529H}).  The mass of the
candidate can span a large range depending on the model. For this
reason, a signal is generally searched for in X-rays
\cite{2009arXiv0911.1774B,2010MNRAS.407.1188B}, or in $\gamma$-rays
\cite{2007JCAP...11..003B}. In the $\gamma$-ray regime, the most
stringent limit on the DM lifetime (that arises in the spectral term)
comes from the non-detection by the Fermi-LAT collaboration
\cite{2009ApJ...697.1071A} of emission from the directions of galaxy
clusters \cite{2010JCAP...12..015D}. 

Here we focus on the astrophysical factor
\begin{equation}
D(l,b,\Delta\Omega)=\int_{\Delta\Omega}\int \rho (l',\Omega) \,dl'd\Omega\;,
\end{equation}
which is the integral of the DM density, $\rho(l',\Omega)$, over line of
sight $l'$ and solid angle $\Delta\Omega = 2\pi\cdot(1-\cos(\alpha_{\rm
int}))$, where $\alpha_{\rm int}$ is the integration angle. When D is computed over
cosmological distances, the spatial term becomes coupled to the energy-dependent term 
as $\gamma$-rays are absorbed along the line of sight ($e^{-\tau(E,z)}$ attenuation 
factor, see, e.g., \cite{2010NuPhB.840..284C}). `Cosmological' $\gamma$-rays will also be
redshifted, affecting the spectrum. Here however, our analysis relies on the MCXC
catalogue of galaxy clusters \cite{piffaretti11}, the redshift distribution of
which peaks at $z\sim 0.1$ (see their Fig. 1) so that we can safely neglect
the above processes. This allows us to factor out energy-dependent effects, i.e. the spectrum resulting
from a prompt emission and also an inverse Compton (IC) contribution from
scattered DM-induced electrons and positrons. The benefit is twofolds:
i) it simplifies the discussion, and ii) also avoids
introducing strongly DM model-dependent factors. Note however, that the
respective IC contributions in the Galaxy and in the clusters are probably
different, and this should be kept in mind when comparing the Galactic
DM diffuse and galaxy cluster $D$-factors.

As in the case of searches for annihilating DM, interesting targets include
dwarf spheroidal galaxies, our own Galaxy, and clusters of galaxies
\cite{2006PhRvL..97z1302B}. Recently, diffuse $\gamma$-ray emission
\cite{2010NuPhB.840..284C} and cross-constraints with other channels (e.g.,
anti-protons and positrons) have also been considered 
\cite{2010JCAP...06..027Z,2010JCAP...07..008H,2011JCAP...09..007C,2011PhLB..698...44K,2011JCAP...01..032G}.
These studies rely on the analysis of single objects or small samples 
constituted by what are thought to be the best targets.  As underlined in
\cite{2011arXiv1110.1529H}, one advantage of combining several galaxy clusters
is that uncertainties (for example on the total mass and DM concentration) somewhat cancel out. 

With this in mind, we show below how the astrophysical term $D$ can be
increased by a stacking strategy of galaxy clusters without compromising
sensitivity due to increasing the amount of background integrated. For that
purpose, we take advantage of the recently published Meta-Catalog of X-ray
detected Clusters, MCXC \cite{piffaretti11}, which encompasses 1743 
clusters of galaxies (making it the largest sample of clusters with detected
X-ray emission from their hot gas).

\section{Method}

 \subsection{Galactic and cluster signal}

All calculations of the $D$-factor are performed with the public code {\it
CLUMPY} v2011.09 \cite{2011_CPC}. The Galactic DM halo is chosen to follow
an Einasto profile \cite{2008MNRAS.391.1685S} with a local density
$\rho_\odot=0.3$~GeV~cm$^{-3}$.  We assume that the DM profiles of all the
galaxy clusters in the catalogue follow a NFW parametrisation
\cite{navarro97}.  For a cluster, we use the standard definition for
$R_\Delta$ to be the radius within which the average density reaches
$\Delta$ times the critical density of the Universe (at a given redshift).
The mass $M_\Delta$ is then simply the mass enclosed within $R_\Delta$. Most
observational constraints and predictions are expressed in terms of
$\Delta=500$ or $\Delta=200$. Using values for $M_{500}$ (derived from the
X-ray luminosity) provided in the MCXC meta-catalogue \cite{piffaretti11}, we
use a mass-concentration relationship \cite{2008MNRAS.390L..64D} to derive
the scale radius and normalisation of the NFW profile for each cluster. To
first order, the uncertainties related to X-ray observations (i.e. $\sim
15$\% average on $M_{500}$), and the scatter related to the
mass-concentration relationship \cite{2008MNRAS.390L..64D} average out if
the stacked signal from a large sample of clusters is considered.

 \subsection{Extragalactic isotropic signal}
As shown in \cite{2010NuPhB.840..284C}, contrary to the DM
annihilation case, the extra-galactic DM decay contribution is a robust
quantity depending mostly on the mean DM density of the universe 
$\Omega_{\rm DM} \rho_{c,0}$. For the sake of simplicity, we neglect the
absorption $\tau(E,z)$ as it has only a moderate impact below 100 GeV
\cite{2010NuPhB.840..284C}. Doing so yields an upper limit for this 
contribution that is written as
\begin{equation}
\label{eq:extragal}
  D^{\rm u.l.}_{\rm extra-gal.}(\Delta\Omega,\, >z)= 
     \Delta\Omega
  \int_z^\infty \frac{\Omega_{\rm DM} \rho_{c,0}}{H(z')}dz'\,,\!\!\!
\end{equation}
where $H$ is the Hubble constant for the concordant $\Lambda$-CDM cosmology.
This quantity is used below for illustration purposes only, as a limit to the continuum signal which can
drown the cluster signal (see in particular Section~\ref{sec: stacking}).

\section{Results}

\begin{figure}[!t]
\includegraphics[width=\columnwidth]{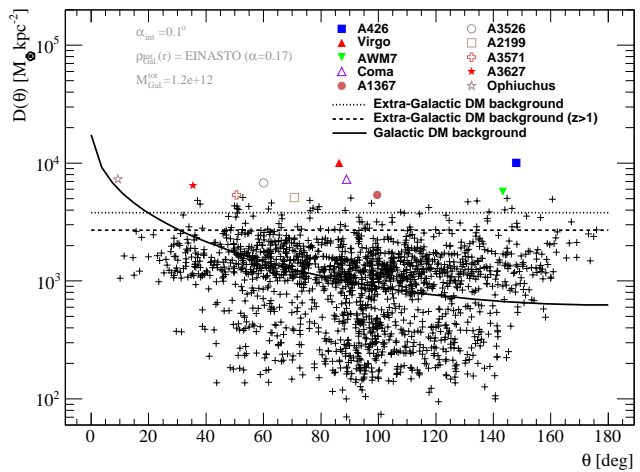}
\caption{Astrophysical factor for decaying DM. Symbols are for galaxy
clusters, solid line is for the Galactic DM background, and the dotted
and dashed lines are an upper-limit (no absorption on the extragalactic
background light) for the isotropic extragalactic DM background (see
Eq.~\ref{eq:extragal}). The 10 brightest objects are identified with
coloured symbols.}
\label{fig:fig1}
\end{figure}

Figure~\ref{fig:fig1} shows the astrophysical term $D$ (in
$M_\odot$~kpc$^{-2}$) for each cluster (plus symbols) as a function of its
angle $\theta$ w.r.t. the Galactic centre direction. A large number of
objects outshine the Galactic DM background (the solid line), fewer
outshine the extragalactic signal upper limit (dotted line). Being a
compilation of various X-ray cluster catalogues, the MCXC meta-catalogue is
not complete at any redshift or in any mass range. This diversity is
reflected by the less populated region below the Galactic signal (solid
line) in Fig.~\ref{fig:fig1}. In principle, an integration along $z>0$
of the extragalactic signal amounts to some double counting of the MCXC
galaxy clusters. The dashed line calculated for $z>1$ illustrates how much
this double counting could be at most (since the MCXC catalogue is not
complete up to this redshift). 

An integration angle of $\alpha_{\rm int}=0.1^\circ$ is adopted,
corresponding to the typical angular resolution of current $\gamma$-ray
instruments well above threshold (e.g., Fermi-LAT for energies above $\sim
10$\,GeV, H.E.S.S. above $\sim$300\,GeV). This choice is appropriate for the
signal from cluster halos, as their typical angular scale (i.e., $R_{500}$)
is $0.1^\circ-1^\circ$. However, we underline that the Galactic and
extragalactic signals scale with $\alpha_{\rm int}^2$, so that smaller
integration angles are in principle favoured to increase the contrast
between the cluster signal and these DM backgrounds.

\subsection{$\log N - \log D$}

\begin{figure}[t!]
\begin{center}
\includegraphics[width=0.55\columnwidth]{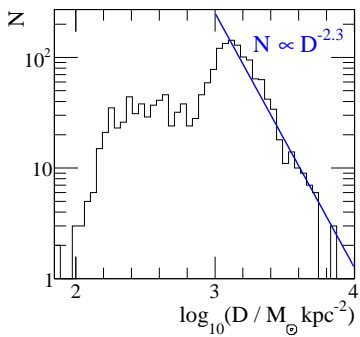}
\caption{$\log N-\log D$ for the galaxy cluster population for $\alpha_{\rm
int}=0.1^\circ$. The behaviour for the brightest objects is consistent with
$N\propto D^{-2.3}$.}
\label{fig:fig2}
\end{center}
\end{figure}

The promise of a stacking analysis for decay is suggested from
Fig.~\ref{fig:fig2}. The $\log N-\log D$ plot has a slope of $-2.3$ at large
values of $D$: the number of objects increases faster than the signal
decreases. The incompleteness of the MCXC meta-catalogue is seen as a drop for
$D<10^3$~$M_\odot$~kpc$^{-2}$. A larger and more complete catalogue  with a
well defined selection function could further increase the gain provided by
a stacking approach if the $N\propto D^{-2.3}$ behaviour continues down to
smaller $D$ values. In a few years from now, the eROSITA mission
\cite{2011SPIE.8145E.247P} should provide such a catalogue. There is naturally a limit to this gain, as at large redshift, clusters may not
have formed yet. The maximum gain can in principle be estimated from the theoretical
redshift distribution of clusters, or from the use of more complete, e.g. optical, 
galaxy cluster catalogues (although these observations poorly constrain the cluster DM content).
For the current study we focus on X-ray identified clusters to reduce uncertainties 
in the DM signal and to provide a list of target objects to current observatories.

\subsection{Stacking}
\label{sec: stacking}

\begin{figure}[t!] \includegraphics[width=\columnwidth]{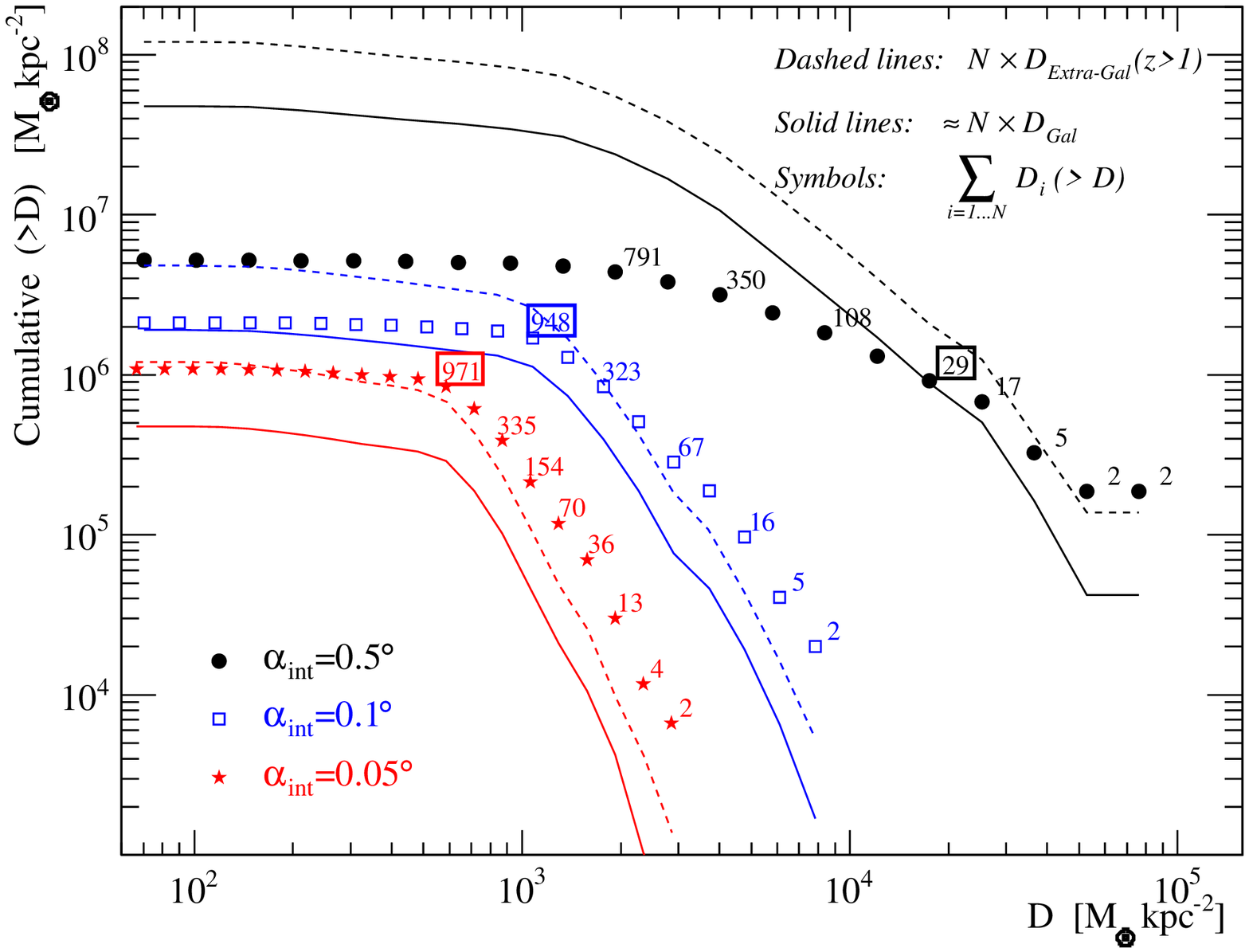}
\includegraphics[width=\columnwidth]{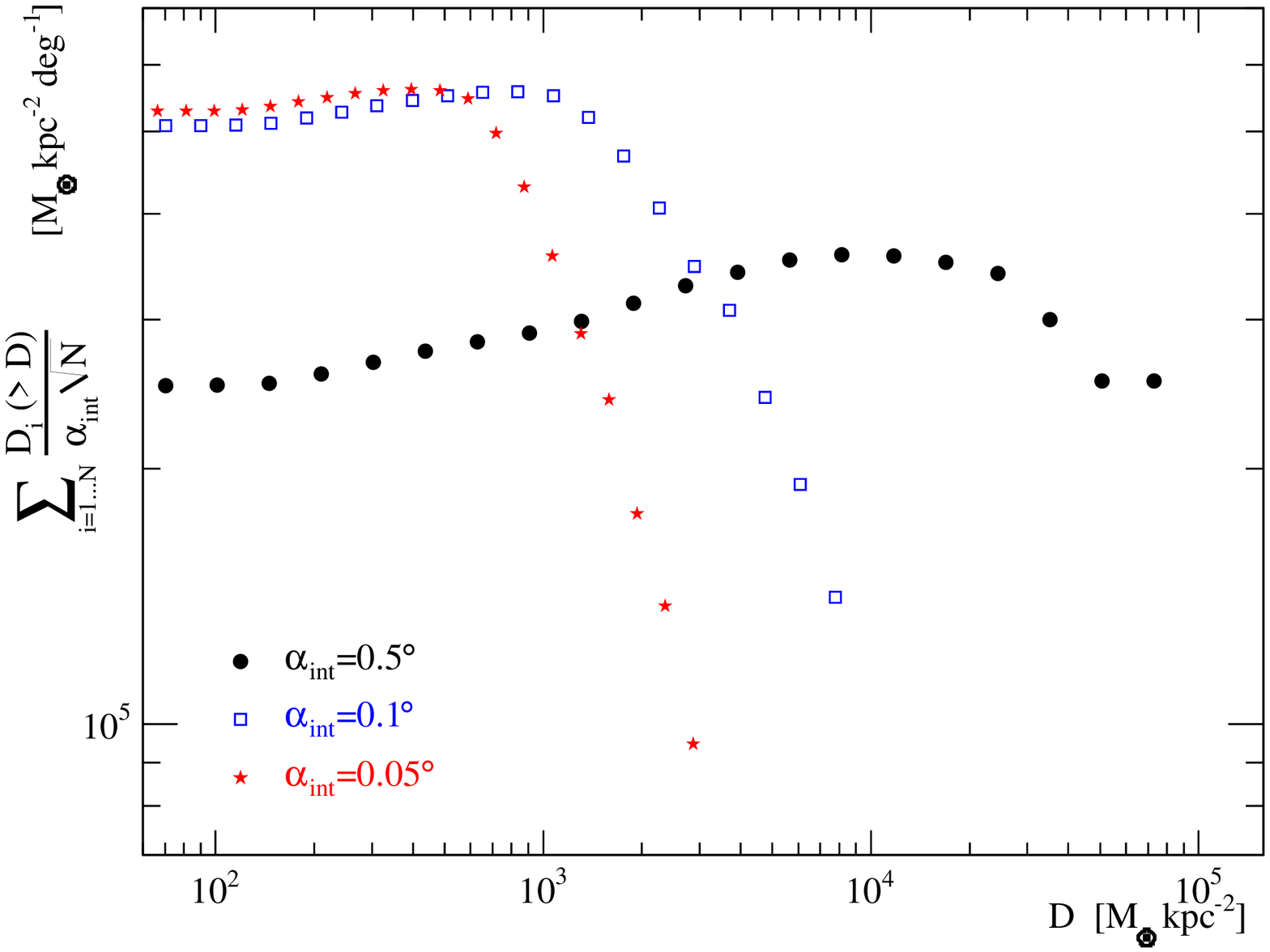} \caption{{\bf Upper panel}:
cumulative galaxy cluster signal (solid symbols) and Galactic (solid lines)
and extra-galactic upper limit (dashed lines) DM background as a
function of $D$. The three sets of curves correspond to three different
integration angles: $\alpha_{\rm int}=0.5^\circ$ (black circles),
$\alpha_{\rm int}=0.1^\circ$ (blue squares), and $\alpha_{\rm
int}=0.05^\circ$ (red stars). The number above a given bin corresponds to
the total number of clusters with a signal larger than $D$. The far
left-hand bins contain a total of 1743 clusters. {\bf Bottom panel}:
cumulative $D$ divided by $\alpha_{\rm int} \sqrt{N}$, proportional to the
signal-to-noise ratio for a given cluster sample.} \label{fig:fig3}
\end{figure}

Figure~\ref{fig:fig3} (upper panel) shows the cumulated signal of all galaxy
clusters having an astrophysical factor $>D$ as a function of $D$.  In all
cases the cumulative signal saturates at $<800$ clusters, consistent with
the drop-out seen in Fig.~\ref{fig:fig2}. For the case of $\alpha_{\rm
int}=0.5^\circ$ (black circles) the integrated $D$ for 791 clusters is
$6\cdot 10^6$~$M_\odot$~kpc$^{-2}$, a factor $\sim 30$ above the value for
the top two clusters (furthest right point). For each $D$ bin, the
corresponding cumulative background signal from the Galactic DM halo 
(resp. extragalactic) is also shown  with solid lines (resp. dashed lines).
The region between the two lines provides the range of accumulated diffuse DM
signal to expect. For 29 clusters, the accumulated background signal from
the Galactic DM halo reaches the level of the cluster signal. The
integration angle of $0.5^\circ$, as used in Fermi-LAT analyses of dwarf
spheroidal Galaxies \cite{2010JCAP...05..025A,2011PhRvL.107x1302A},
corresponds to the Fermi-LAT angular resolution at around 1~GeV, degrading
to $\sim 1^\circ$ at lower energies and reaching $0.1^\circ$ above 10~GeV.
The blue squares and red stars show the result of repeating the analysis for
$\alpha_{\rm int}=0.1^\circ$ and $0.05^\circ$ respectively (a resolution
generally within the capabilities of X-ray instruments, or within reach of
the next generation of air-Cerenkov $\gamma$-ray  telescopes). In these
cases the accumulated Galactic DM plus extra-galactic signal is almost 
always below the cumulative cluster signal. Because of a smaller
integration angle, saturation is not reached at the same cumulated value.
For $\alpha_{\rm int}=0.1^\circ$ a gain of a factor $\sim100$ in signal is
observed between stacking two and 948 sources in the cumulative $D$.  

The best observational strategy depends not only on the integrated signal
and Galactic halo plus extra-galactic DM background, but on the
(generally much higher) level of charged particle and diffuse astrophysical
gamma-ray backgrounds. In the background-limited regime the best approach is
to maximise the signal-to-noise ratio or signal divided by the square-root
of background. In the case of uniform background the signal-to-noise is
proportional to the cumulative $D$ divided by $\alpha_{\rm int} \sqrt{N}$
(where $N$ is the number of clusters with $D_{\rm i}>D$), as shown in the
bottom panel of fig.~\ref{fig:fig3}. The peaks in these curves (at 29, 948
and 971 clusters for $\alpha_{\rm int}=0.5^\circ$, $0.1^\circ$ and
$0.05^\circ$ respectively) correspond to the optimum stacking approach in
the background-limited, rather than signal-limited, regime. With the current
MCXC meta-catalogue,  there seems to be no advantage in resolutions better
than $0.1^\circ$. In the signal-limited regime a larger integration angle is
preferred. Stacking leads to a signal-to-noise increase of a factor $\sim5$.
With a more complete galaxy cluster catalogue, the `saturation' regime (the
break seen in the upper-panel curves) would be moved towards smaller $D$
values and the benefit of the stacking approach could be greater. 

\section{Discussion and conclusion}

As well as (close to uniform) charged particle backgrounds and the diffuse
Galactic DM and astrophysical backgrounds (see e.g. 
\cite{2011PhRvL.107x1302A}), non-DM gamma-ray emission from galaxy clusters
must be considered as another possible source of background and/or confusion
\cite[see][]{2009PhRvD..80b3005J,2011arXiv1105.3240P}.  No individual galaxy
cluster -- but for Virgo, for which a gamma-ray emission was detected
positionally consistent with the centre of M87 by Fermi-LAT
\cite{2009ApJ...707...55A} -- has been detected in gamma-rays so far, but
accelerated particles are expected to provide an additional gamma-ray signal
in clusters.  Given the present situation of non-detection in the direction
of individual massive halos, a stacking analysis is advantageous to place
limits on DM decay.  Any object for which a signal (of suspected non-DM
origin) is observed can be discarded in the analysis at little cost for the
exclusion limit. On the other hand, if any signal is observed and is
consistent with an astrophysical origin, stacking or merely looking
at galaxy clusters will be extremely challenging for DM searches. Alternatively, the
strong scaling $\log N-\log D$ could be a good diagnostic to disentangle a
decay from an astrophysical signal if the latter shows a different scaling
(Maurin et al., submitted). The case of annihilating DM, which is found to
be less favourable but more crucial physically (limits of non-detection are
close to the cross-section expected from that required to match the thermal
relic abundance) is discussed in Nezri et al. (in preparation), focusing on
the instrumental response of current and future observatories (Fermi and
CTA).

To ease further analysis based on our model, we provide with this paper a
file containing  all the necessary inputs for each cluster. 
An analysis on real data can certainly be optimised by adapting the integration
region for each cluster. Assuming all galaxy clusters share the same DM
profile, and given the mass range spanned by the MCXC, we can
make the first order approximation that their concentration
parameters are the same. In other words, the
concentration for an NFW profile is $c(M) = R_{\rm vir}/r_s$ and we assume
$c(10^{14} M_\odot)\sim5$ \cite{2008MNRAS.390L..64D}. Defining
\begin{eqnarray}
  \alpha_s    &\equiv& \tan^{-1}\left(\frac{r_s}{d}\right),
     \,\,\, \alpha_{\rm max}   \equiv \tan^{-1}\left(\frac{5r_s}{d}\right),
     \nonumber\\
  \label{eq:alphas}
  x           &\equiv& \frac{\alpha_{\rm int}}{\alpha_s}, \,\,\,
  {\rm ~~and~~~} x_{\rm max} \equiv \frac{\alpha_{\rm max}}{\alpha_s}\approx 5,
    \nonumber
  \label{eq:xmax}
\end{eqnarray}
we find that there is a universal dependence of the fraction 
\begin{equation}
{\cal F}_D(x) \equiv \frac{D(x\cdot\alpha_s)}{D_{\rm max}}
\end{equation}
on the DM-decay signal. Hence, given $D(0.1)^\circ$ and $\alpha_s$ (available in
the supplementary file \footnote{See Supplemental Material at {\tt[enter URL]} for the D-values of all MCXC objects.}, a sample of which being given in table~\ref{tab:tab1}), and a parametrisation of this universal dependence
\begin{equation}
 {\cal F}_D(x) \!\!=\!\!
    \begin{cases}
      e^{\left[-1.17 + 1.06  \ln(x)
           - 0.17 \ln^2(x) -0.015 \ln^3(x)\right]} \text{~~~if $x \leq 5$,}\\
       1  ~~~~~\text{otherwise;}
    \end{cases}
\end{equation}
the term $D$ can be calculated for any integration angle, using
\begin{equation}
 D(\alpha_{\rm int})=  D(0.1^\circ) \times
   \frac{{\cal F}_D \left(\alpha_{\rm int}/\alpha_s\right)}{
   {\cal F}_D \left(0.1^\circ/\alpha_s\right)}\,.
\end{equation}
This parametrisation describing the fraction of the signal in a given angular
region is valid down to ${\cal F}_D=10^{-3}$. We note that this
function is unchanged if an Einasto profile is used, but the D-values
obtained from such a profile are larger than that calculated with an
NFW DM distribution.

The values for the ten brightest clusters are given in Table~\ref{tab:tab1},
with values of $D$ provided for a $0.1^\circ$ integration angle (see
Supplemental Material [37] for the full list of clusters). We have repeated (not
shown) the calculation for $\alpha_{\rm int}=1^\circ$ (an angle for which
most clusters fall below the Galactic DM background) to compare the $D$
factors with those of \cite{2010JCAP...12..015D,2011arXiv1110.1529H}: our
values for Fornax, Coma, AWM\,7 and NGC\,4636  are found to be lower, but
within a factor 2 of what was found in these previous works. In
particular, the large difference for the Fornax cluster explains why these
authors have flagged it as being the best target for DM decay when it does
not make it in our top-ten. This is understood as follows. Decay is very
sensitive to the mass estimate (which goes directly as the integration of
the density). In \cite{2010JCAP...12..015D,2011arXiv1110.1529H}, the authors
use $M_{500}$ values from the HIFLUGCS catalogue 
\cite{2002ApJ...567..716R,2007A&A...466..805C} that are larger than the ones
provided in the MCXC catalogue (e.g., factor of $\sim$5, 2, and 2 for Fornax,
Coma, and AMW7 respectively). For the former analyses, the authors used a $\beta$-model
for the gas distribution
\cite{2002ApJ...567..716R,2007A&A...466..805C}, whereas the MCXC relies on
the more accurate AB-model \cite{2002A&A...394..375P}. As discussed in the
App.~A of \cite{piffaretti11}, beta-models can produce factor 2 higher, or lower, 
values of M500, depending on the core radii used. The masses for the MCXC 
sample could still suffer from systematics, 
as for instance from the use of the hydrostatic equilibrium hypothesis. 
However many comparisons to numerical simulations indicate that these 
uncertainties are $<15-20$\% \cite{2008A&A...491...71P}. In any case, 
if the signal from the brightest source is overestimated (resp. underestimated) by a factor $f$
the improvement factor will increase (resp. decrease) roughly by the same
amount. For the MCXC, we recall that $M_{500}$ and $R_{500}$ are
obtained from the X-ray luminosity, making use of the $L_X-M_{500}$ scaling
relation  derived for the REXCESS sample by \cite{pratt09}. The mass
estimate not only links to the $L_X$ measurement, but also to the intrinsic
scatter about the scaling relation. Further conversions to $R_{200}$ may
amplify this dispersion effect. This stresses that caution is advised when
working with individual clusters for which mass and radius are inferred from
the X-ray luminosity. In that respect, stacking provides a more robust
approach as it washes out those uncertainties.

\begin{table}
\begin{center}
\caption{Ten brightest galaxy clusters (positions taken from the MCXC
meta-catalogue \cite{piffaretti11}) in DM-decay for $\alpha_{\rm
int}=0.1^\circ$\vspace{-0.05cm}.}
\label{tab:tab1}
\begin{tabular}{@{}lccccccccr} 
\hline\hline 
Name     &$\!\!$Index$\!\!$&~~&  $l$  &  $b$  &  $d$   &~~&$\alpha_s$& $D(0.1^\circ)$\\
&$\!\!$(MCXC)$\!\!$&~~& (deg) & (deg) &  (Mpc) &~~&   (deg)  & ($M_\odot~{\rm kpc}^{-2})\!\!$ \vspace{0.05cm}\\ \hline
A426       &258&~~& 150.6 & -13.3 &  75.0  &~~&   0.44   & $1.0 \cdot 10^{4}$ \vspace{0.0cm}\\
Virgo      &884&~~& 283.8 &  74.4 &  15.4  &~~&   1.09   & $1.0 \cdot 10^{4}$ \vspace{0.0cm}\\
Coma       &943&~~&  57.2 &  88.0 &  96.2  &~~&   0.29   & $7.3 \cdot 10^{3}$ \vspace{0.0cm}\\
Ophiuchus$\!\!\!$&1304&~~&   0.6 &   9.3 &  116.  &~~&   0.27   & $7.3 \cdot 10^{3}$ \vspace{0.0cm}\\
A3526      &915&~~& 302.4 &  21.6 &  48.1  &~~&   0.39   & $6.8 \cdot10^{3}$ \vspace{0.0cm}\\
A3627      &1231&~~& 325.3 &  -7.1 &  66.0  &~~&  0.32   & $6.5 \cdot 10^{3}$ \vspace{0.0cm}\\
AWM7       &224&~~& 146.3 & -15.6 &  72.1  &~~&   0.28   & $5.7 \cdot 10^{3}$ \vspace{0.0cm}\\
A1367      &792&~~& 235.1 &  73.0 &  89.3  &~~&   0.24   & $5.4 \cdot 10^{3}$ \vspace{0.0cm}\\
A3571      &1048&~~& 316.3 &  28.6 &  160.  &~~&  0.18   & $5.4 \cdot 10^{3}$ \vspace{0.0cm}\\
A2199      &1249&~~&  62.9 &  43.7 &  124.  &~~&   0.20   & $5.1 \cdot 10^{3}$ \vspace{0.0cm}\\ \hline
\end{tabular}
\end{center}
\end{table}

To conclude, we have shown that a stacking approach can help to push down by
a factor of  5 (for the background-limited regime) to 100 (in the
signal-limited case) the current observational limits on decaying DM (when
using the MCXC meta-catalogue of galaxy clusters). A more thorough
analysis (e.g., using deeper galaxy cluster catalogues and including
absorption on the extragalactic background, uncertainties on DM profiles,
etc.) is left for future work. Meanwhile, we provide the necessary inputs
to apply this idea to existing experiments (for example Fermi-LAT).

\

\acknowledgments
We thank the anonymous referee for the constructive criticism
that helped clarify the paper. RW is supported by an STFC post-doctoral fellowship.


\end{document}